\documentclass[conference]{IEEEtran}
\usepackage{cite}
\usepackage{amsmath,amssymb,amsfonts}
\newtheorem{example}{Example}[section]
\newtheorem{theorem}{Theorem}[section]
\usepackage{graphicx}
\usepackage{textcomp}
\usepackage{xcolor}
\usepackage{makecell}
\usepackage{hyperref}
\usepackage{paralist}
\usepackage{fullpage}
\usepackage{times}
\usepackage{fancyhdr,graphicx,amsmath}
\usepackage[ruled,vlined]{algorithm2e}
\usepackage{subcaption}
\usepackage{bm}

\renewenvironment{thebibliography}[1]{%
\begin{oldthebibliography}{#1}%
    \setlength{\parskip}{0\baselineskip}
    \setlength{\itemsep}{0\baselineskip}
}%
{%
\end{oldthebibliography}%
}
\def\BibTeX{{\rm B\kern-.05em{\sc i\kern-.025em b}\kern-.08em
    T\kern-.1667em\lower.7ex\hbox{E}\kern-.125emX}}
\begin{document}

\title{Experience-Enhanced Learning: One Size Still does not Fit All in Automatic Database Management.}

\author{\IEEEauthorblockN{Yu Yan,Hongzhi Wang, Jian Ma, Jian Geng, Yuzhuo Wang}
\IEEEauthorblockA{\textit{School of Computer Science and Technology} \\
\textit{Harbin Institute of Technology}\\
Harbin, China \\
yuyan@hit.edu.cn,wangzh@hit.edu.cn,1253791041@qq.com,1161906117@qq.com,wangyuzh@hit.edu.cn}

}
\vspace{-5mm}
\maketitle

\begin{abstract}
  Recent years, the database committee has attempted to develop automatic database management systems. Although some researches show that the applying AI to data management is a significant and promising direction, there still exists many problems in implementing these techniques to real applications (\textbf{long training time, various environments and unstable performance}). In this paper, we discover that traditional rule based methods have the potential to solve the above problems. We propose three methodologies for improving learned methods, i.e. \textbf{label collection for efficiently pre-training, knowledge base for model transfer and theoretical guarantee for stable performance}. We implement our methodologies on two widely used learning approaches, deep learning and reinforcement learning. Firstly, the novel experience enhanced deep learning (EEDL) could achieve efficient training and stable performance. We evaluate EEDL with cardinality estimation, an essential database management. The experimental results on four real dataset~\cite{ready} show that our EEDL could outperforms the general DL model~\cite{mscn}. Secondly, we design a novel experience-enhanced reinforcement learning (EERL), which could efficiently converge and has better performance than general RL models~\cite{reindex}. We test EERL with online index tuning task. The experiments on TPC-H shows that EERL could accelerate the convergence of agent and generate better solution that generalizes the reinforcement learning. 
\end{abstract}

\section{Introduction}
\label{introduction}
    With the development of database systems, the heavy manual cost and the complexity of database management tasks bring the great demand of automatic database management.
On the one hand, the expenses for Database Administrators (DBAs) became the increasingly important part in the Total Cost Ownership (TCO) with the advent of Database as a Service (DaaS)\cite{cost}. Database vendors urgently need the automatic solutions to reduce labor costs. On the other hand, complex environment of database including database engine, software systems and hardware conduct difficult database management problems\cite{complex}, such as the index tuning of cloud database, the knob configuration of distribution database and etc. It is difficult for DBAs to understand such complex environment or adapt to the change environment in time. As a result, only relying on the intuition of DBAs may conduct some poor performance.

\begin{table*}[h]
    \centering
    \begin{tabular}{||c c c c c c||}
        \hline
            Type  & Relational & Key-value & Document & Time Series & Graph \\
            \hline \hline
            Number & 147 & 57 & 52 & 34 & 34 \\
        \hline
    \end{tabular}
    \caption{The number of each type database engine (from DB-Engines October 2021).}
    \label{database}

\end{table*}

Recently, researchers have proposed many learned methods~\cite{automaticM,databaseT,card1,queryO1} for automatic database management. These solutions have produced some inspired results in many database management areas, such as configuration tuner~\cite{automaticM,databaseT}, cardinality estimation\cite{card1, card2} and query optimization\cite{queryO1,queryO2}. Importantly, some learned database management methods perform better than the existing rule-based solutions, and has the potential to replace traditional ones\cite{survey}. Thus, some popular researches show that applying machine learning to database management is a promising direction\cite{automaticM,databaseT,card1,queryO1}

However, there still exists significant problems in deploying these methods in practise. From existing learned database management approaches\cite{automaticM,databaseT,card1,queryO1,reindex,endtoend}, we conclude three issues which render them implementing in practice, i.e.(i) \textbf{long training time},(ii) \textbf{various environments} and (iii) \textbf{unstable performance}.

(i)Almost all learned methods in database management system require a large amount of high quality labeled data for model training. For example, the end-to-end cardinality estimator\cite{endtoend} based on deep learning generates 10K queries for model training. Even the unsupervised reinforcement learning method\cite{reindex} also requires at least thousands of interactions with the real environment for obtaining the reward value before they could be used, which is also cost much time in query execution. However, different with other area such as natural language processing and computer vision, due to the complexity of database, obtaining labels for database management tasks will take much time, which could not be accelerated by GPU\cite{2020Magic}. Especially, obtaining labels in massive data takes a large time cost that users could not afford for model training. 

(ii) Database environment has a great variety. For example, as shown in Table~\ref{database}, only the relational database has 147 kinds of database engines.  Especially, for one database engine, the difference versions also have significant difference.  The original version of MySQL is completely different with the new version. 
Various environments have different requirements for database management models. Thus, the models trained in a certain situation may be unsuitable for other situations. As a consequence, we could not use one model to fit all situations, model should have the ability to adjust to environments.

(iii) Different from the rule-based database management solution, the learned model is a black box. On the one hand, the performance of learned methods is unstable with the different parameters initialization, training algorithm, iteration times and activate functions. There may conduct different results with even the same model and environment. On the other hand, the data distribution of the learning task is vital to the performance of the learning results. Especially, the irregular data or query distribution in database management systems may lead to poor performance. For example, in query optimization, there always exist some outlier queries which lead to the tail catastrophe\cite{queryO2}. By contrast, the high availability of database management system requires a relatively robust solution.  Thus, in database management system, we should ensure the availability of the approaches.

From above discussions, we can find that existing learned approaches have vital problems which make them unsuitable in real applications. This motivates us to find solutions for these problems. Completely different with the learned methods, there are another type of statistical methods\cite{cophy,greedy} based on rules which could also implement automatic database management. These rule-based methods are more stable but poorer in performance than learned ones\cite{neo}. Thus, we attempt to involve rule-based methods to learned ones to solve their problems in the following three aspects.

(i) Almost all the experience rule-based solutions\cite{drop,genetic,relax,cophy,mhist,quick} mainly rely on the experiences, there no need for much more time consumption in the preparation stage. For example, the traditional index selection algorithm\cite{relax} achieves index recommendation with five important experience rules, i.e. merge, split, prefix, promote, remove, which are concluded from practices. Such method could be directly implemented in relational cases without training. Therefore, the rule-based method may have the potential to accelerate the training process of learned methods.

(ii) Rule-based solutions\cite{genetic,mhist} do not consider a ultimate optimization for processing certain situation, these rules are concluded to generally solve a certain management task. Therefore, these methods usually achieve better universality than learned methods. For example, the cardinality estimator in MySQL is unified and could process all the relational schemas, while an estimator based on deep learning\cite{card1} is a unique model and could only process a certain data\&workload instance. Thus, embedding rule-based methods to learned ones may make them more general.

(iii)Compared with learned methods, rule-based methods have more solid theoretical guarantee\cite{greedy}.
These methods could guarantee the more stable performance of database management task. For example, the index selection algorithm\cite{cophy} which is based on the linear programming algorithm could stably find the index solution with any space budget. As a consequence, rule-based methods have the potential ability to help the learned method to guarantee a stable performance for the high availability of database management system.

Based on above reasons, we propose rule-enhanced learned methods in this paper. Aiming at the three problems of learned methods, we design three methodologies which integrates rule-based methods to enhance the learned methods. For reducing time consumption of model training, our key idea is to minimize the time consumption in query execution, and we propose the experience-enhanced label collection (ELC) method which sacrifice the label accuracy for time cost. For the adaptation to various environments, we design an experience knowledge base (EKB) which utilizes the rule-based methods as the transition to help learned methods to achieve stable transfer learning. For stable performance, we design a solution evaluation algorithm (SEA) which could guarantee a theoretical bound for learned solution by controlling the credibility between the learned solution and the rule-based solution.

Moreover, we implement our methodologies on two popular learned methods, deep learning and reinforcement learning, which are widely used in learned database management. For deep learning, we utilize ELC to accelerate the model training, integrates EKB to help transfer learning and employ the SEA to support the stability of the learned solutions. Compared general learned methods, EEDL has shorter training time and better adaptation for various environments. 
For reinforcement learning, we utilize the EKB to accelerate the adaptation to environment and employ the ELC to efficiently obtain reward value. Compared to general RL, EERL could quickly converge and has more stable performance. Our contributions are summarized as follows:s

\begin{itemize}
    \item We firstly observe that rule-based methods have the potential to solve the problems of learned methods. We propose three novel rule-embedding methodologies to solve the challenges of deployment learned methods. To our best knowledge, this is the first one to solve these problems by enhancing the experience rule-based methods. 
    \item We implement our methodologies to the deep learning approaches and design a rule-enhanced deep learning approach, EEDL. After some evaluation in cardinality estimation task, we find that EEDL could reduce the time consumption and improve management precision. Specifically, our method could accomplish model training with several seconds in open source dataset~\cite{census,dmv} and achieve higher precision than general learned method~\cite{mscn}.
    \item Except the deep learning, we also implement our methodologies to another popular learned method, reinforcement learning. We design a novel reinforcement learning methods, EERL. The experimental results in index tuning task clarify that EERL could quickly converge and has more stable performance than general RL\cite{reindex} in open source benchmark TPC-H. 
\end{itemize}

Supplementally, our paper clarifies a novel development direction for learned solutions, which claim that traditional rules could enhance the learned methods. We hope that our paper could spark more promising practical results.
The remainder of this paper is organized as follows: In Section~\ref{related}, we conclude the existing automatic database management methods. We present our methodologies in Section~\ref{methods}, which use experience rule-based methods to enhance the learned methods. In Section~\ref{eedl}, we introduce the detailed implementation of our methodologies in deep learning. And in Section~\ref{evaeedl}, we show the experimental results for EEDL. In Section~\ref{eerl}, we present the experience enhanced reinforcement learning. And in Section~\ref{evaeerl}, we introduce the experimental result of EERL with online index tuning task.

    \section{Related Works}
    \label{related}
        
With the development of database, the automatic database management methods become increasingly important. In this section, we conclude some related automatic database management methods, including knob tuning, index selection, view selection and cardinality estimation. Next, we introduced some typical works of above aspects.

\subsection{knob tuning}

Knob tuning is an important task of database system, which directly influence the efficiency of database~\cite{2019Speedup}. Existing tuning methods consist of the experience rule based~\cite{2017BestConfig} learned methods~\cite{2019An} and etc. Zhu et al.~\cite{2017BestConfig} uses a recursive searching method to find the best knob, which has more stable performance. And Zhang et al.~\cite{2019An} propose a reinforcement learning model to achieve knob tuning which has an ultimate tuning for current environment.

\subsection{index selection}

Index is the key part for accelerating query execution in database management system. The index selection refers to find the best indexes which have high time efficiency with lower storage cost. There are two main kinds, offline index selection~\cite{cophy,reindex,1997An} and online index selection~\cite{shift,onlineindex}. And there also exists experience rule-based one and learned one. Dash et al.~\cite{cophy} presents a linear programming methods for index selection, which could stably find an optimal solution. Lan~\cite{reindex} design a reinforcement learning structure for offline index selection, which is unstable to environment changes.

\subsection{view selection}

View improves the query execution by establishing the query result store. This method utilize the space-time trade-off principle~\cite{2015Robust,view}. Dokeroglu et al.~\cite{2015Robust} integrate some searching algorithm(HillClimbing~\cite{Selinger2013Hill}, Genetic~\cite{geneticsear}) to optimize the view. This method has a solid theoretical bound for view selection. And Yuan et al.~\cite{view} attempt to use the reinforcement learning to select view which has a unique results after some interactions with environments.

\subsection{cardinality estimation}

Cardinality estimation is a essential task in database. Researchers has proposed many methods for this task~\cite{1992Practical,deepcard,card1,card2,card3}. These methods could be divided into two classes, statistics-based and learning-based. Lipton et al.~\cite{1992Practical} propose a cardinality estimation through adaptive sampling, which is based on statistics. And Yang et al. propose an learned estimator to implement the cardinality estimation, which has higher accuracy than traditional ones~\cite{1992Practical,mhist}.

\section{Our methodologies} \label{methods}
    In this section,
we design the methodologies to solve the problems of existing learned methods.
In order to reduce the training time, we propose experience-enhanced label collection(ELC) approach in Section~\ref{sec:label}. ELC achieves efficient model pre-training by using rule-based methods to minimize the time consumption of labeling in query execution, which could hardly be accelerated by GPU. Then, in order to achieve good adaptability for various environments, we design an experience knowledge base(EKB) which includes the rule-based methods to support the effective transfer learning of existing learned methods to the new environment in Section~\ref{know}. To achieve a theoretical guarantee for stable learning, we design a novel solution evaluation algorithm(SEA) which guarantee the stability of learned methods by limiting the credibility between rule-based solution and learned solution in Section~\ref{sec:theore}.

\subsection{Efficient Label Collection}\label{sec:label}

 Efficiently collecting labels can accelerate the model training process for learned database management approaches. In DBMSs, due to the cost of query execution, collecting real labels is very expensive, especially for massive data. In this paper, we propose a methodology which integrates rule-based methods to reduce the time consumption. Our key idea is to use rules to generate the labels that could not be produced according to the system log and have to be generated by taking a long time to run the queries. As shown in Algorithm~\ref{alg:label}, we propose a methodologies(ELC) which uses the rule-based methods to enhance the label collection. Three approaches to complete the label data collection from real-world database, including \textbf{Log Searching}(Line 2-4), \textbf{Manual Generation}(Line 5-8) and \textbf{Waiting Task}(Line 9-12). These three approaches are used in some existing learned works and could cover all situations of label collection of database~\cite{card3,ready,ibtune}. 
\begin{algorithm}
    \SetAlgoLined
    \LinesNumbered
    \SetKwInOut{Input}{input}
    \Input{
        \ $M$ is the management task \\
 
    }
    \SetKwInOut{Output}{output}
    \Output{
        \ $TS$ : the labeled training set
    }

    \BlankLine
    \ $ TS \leftarrow \varnothing $ // \textit{\textcolor{blue}{init the labeled dataset}} \\

        \If(){ hasLog(M) }{
           $TS \leftarrow TS \cup  log.searchLabel(M)$ // \textit{\textcolor{blue}{search labels from database log}} \\
        }

        \If(// \textit{\textcolor{blue}{ TS is not enough}} ){ TS.lack() \&  findRule(M) }{
            \ $MS \leftarrow M.generate() $ // \textit{\textcolor{blue}{generate enough management taks}} \\
            \ $TS \leftarrow rule.generate(MS)$ // \textit{\textcolor{blue}{generate labels by rules}} \\
        }

        \If(// \textit{\textcolor{blue}{still not enough}}){TS.lack()}{
            $replace(learned, rule based)$ // \textit{\textcolor{blue}{use the rule based methods which could cold start}} \\
            $waitingTask(M)$ // \textit{\textcolor{blue}{Wait for the user behaviors}} \\
        }

\caption{Collect Labeled Data for Database Management Task}\label{alg:label}
\ return $TS$ \\
\end{algorithm}

\textbf{Log Searching:}
Since the log contains the information of user operations, it is an efficient solution to gather label set from the log. Some database management tasks can directly dig out enough training data from log, such as cardinality estimation~\cite{queryO2}. We could find the queries and corresponding cardinality from the query history in database log. If we could find enough labeled dataset from log, we do not need to utilize other collection methods.

\textbf{Manual Generation:} For some tasks for which it is difficult to find enough information to construct training set~\cite{reindex}, manual efforts are required. 
Our manual generation method relies on the human to generate task data first.

Specifically, according to the distribution or templates of current task, ELC manually generate enough unlabeled data which cover all situations as soon as possible. Then, ELC utilize the rule-based methods of current task to label, avoiding the cost of query execution. Example~\ref{eg:label} demonstrates an manual label generation for cardinality estimation.

\begin{example}\label{eg:label}
For cardinality estimation, the input consists of query templates predefined by human being, tables and corresponding rules(such as the estimators of MySQL), and the output is the labeled training data set. We first manually generate queries enough to cover all situations by filling these templates with tables. Then, ELC uses the estimator of DBMS to efficiently obtain enough labeled data. Although these labels are not the real cardinality, the labels generated by the rules are reasonable since the rules summary the distribution of the historical data. Therefore, the model trained by these labels could perform well.

\begin{itemize}
    \item Input: query templates, tables.
    \item Output: enough labeled data.
    \item query generation: generate queries according the templates.
    \item  labeling: use the estimator of DBMS to generate labels.
\end{itemize}
\end{example}

\textbf{Waiting Task:} Without the data distribution of the task, we could not generate enough labeled data to cover current task. In such case, the worst solution is to obtain the data gradually through the operations of users. Before the operations of users exist, to enable the system, we use rules to generate some reasonable labels. \textcolor{black}{For example, in index tuning, we could not generate the training data of index plan for the learned model without the features of workload. Then, we need to wait for user actions to collect training data. Because model could not work for queries due to lack of training data. For stable performance, we could use rule-based methods to replace the learned model for index tuning, such as establishing indexes in primary key. }

\subsection{Experience Knowledge Base for transfer Learning}\label{know}

\begin{figure}[htp]
    \centering
      \includegraphics[width=1.0\linewidth]{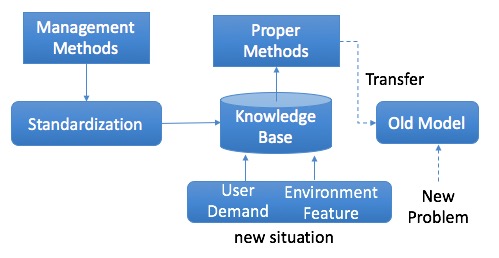}

  \caption{Workflow of Experience knowledge Base}
  \label{fig:know}

  \end{figure}
A significant problem of learned method is that the model trained on one environment may be unsuitable for others~\cite{ready}, while rule-based methods have more general ability~\cite{cophy}. In order to help the effective transfer of the models in learned methods, we design an experience knowledge base (EKB) which integrates some widely used rule-based methods of database management system. A EKB could be formally represented as a triple $(\mathcal{F}(Method), \mathcal{H}(Task, Env), \mathcal{K}$). $\mathcal{K} = \{k_1, k_2,...\}$ is the set of rule-based database management methods. $\mathcal{F}$ is a function to standardize the methods to a unified expression $k_i$. $\mathcal{H}$ is function to find the suitable rule-based method for a certain task and environment. 

Figure~\ref{fig:know} demonstrates the workflow of our methodology. EKB first uses the standardization module to unify some widely used rule-based database management methods. Then, EKB stores these standard methods to knowledge base. Importantly, we develop a schema for the rules in automatic database management $k_i = (input, output, parameter, feature)$. Since each piece of knowledge in the base is a rule-based approach, the schema should include the basic factors of rule-based approaches including the input, output, the parameter of algorithms and a feature vector showing the applicable scenarios. \textcolor{black}{For example, drop~\cite{drop} could be standardized to $k_{index} = ([table, workload], indexes, [stop criterion],$\\
$[goal, multi, relational]) $.}

After the establishment of knowledge base which consists of rule-based method, EKB can use it to accelerate the adaptation of model for changed environments. To find the proper methods in EKB, we vectorize the use demand and environment as $v$. Then, EKB return the suitable rule-based approach $r$ which is most close to $v$. \textcolor{black}{For example, the vector $v$ of index selection $[cost, multi, relational]$ is closer to $k_{index}$ than $[time, multi, graph]$. }

Then, the model transfer consists of two steps. Firstly, EKB is searched to retrieve the suitable approach $r$. The $r$ is transferred to fit the new problem. Since $r$ is a rule-based approach, the transferring is pretty efficient. In the step 2, we gradually gather labeled data by using Algorithm~\ref{alg:label} for implementing model re-training. Our key idea of EKB is to use the rule-based approach as a transition of transfer for learned models, which improve the stability of transfer learning.

\begin{example}\label{eg:know}
   We give an simple example about index selection task to clarify the $\mathcal{K}$. The knowledge base($\mathcal{K}$) consists of the standard rule-based algorithms. For index selection methods, each tuple is informed of (input, output, parameter, solution features), whose semantics are as follows.                                                
    \begin{itemize}
        \item Input: workload, tables.
        \item Output: the optimal indexes
        \item parameter: stop criterion, index width 
        \item Solution Features: optimal goal, multi-columns support, index interaction, online or offline, What-if calls and etc.
    \end{itemize}

\end{example}

\subsection{Theoretical Guarantee for Stability}\label{sec:theore}

 Most learned methods do not have solid theoretical guarantee which may have problems to apply them in high available database system~\cite{survey}. In order to guarantee the stability of the learned methods, we design a methodology (SEA) to support theoretical guarantee for learned methods by calculating the credibility between the learned solution and rule-based solution. The basic idea is that if the the credibility between a rule-based method $C_R$ and a learned method $C_L$ has a bound, and $C_R$ has some performance guarantee $\epsilon$, the performance guarantee of $C_L$ could be derived as $\epsilon$ and the credibility bound.

 \begin{algorithm}
    \SetAlgoLined
    \LinesNumbered
    \SetKwInOut{Input}{input}
    \Input{
        \ $t$ is a database management task \\
        \ $d$ is the bound of credibility\\

    }
    \SetKwInOut{Output}{output}
    \Output{
        \ $C_*$ : the solution of current task t.
    }

    \BlankLine
    \ $C_R = Rule(t) $  // \textit{\textcolor{blue}{solution of the rule-based}} \\
    \ $C_L = Model(t)$ // \textit{\textcolor{blue}{solution of the learned}} \\
    \ $c = \frac{|C_L-C_R|}{C_R}$// \textit{\textcolor{blue}{calculate the confidence }} \\
    \ // \textit{\textcolor{blue}{choose solution}} \\
    \ \uIf{$c < d$}{
             $C_* =  C_L$
        }
    \ \Else(){
        $C_* = C_R$
    }
    return $C_*$

\caption{Choose the Optimal Solution by Credibility}\label{alg:theore}

\end{algorithm}

 The pseudo code is shown in Algorithm~\ref{alg:theore}. Lines 1-2 first obtain two solutions for a certain task, the learned solution $C_L$ and the rule-based solution $C_R$. Then, we calculate the credibility $c$ by utilizing the $C_R$ (Line 3). If $c$ satisfies the given credibility bound $d$, our methodology will return $C_L$ as the optimal solution (Line 6). Otherwise, $C_R$ will be returned (Line 8). Our methods support a stable solution by the credibility guarantee which is calculated according to the rule-based solution. We prove this point as follows.

 We define the solution of learned methods $C_L$, the solution of rule-based methods $C_R$ and the optimal solution $C_*$. And the solution of experience rule-based method has a theoretical bound $ \varepsilon (n)  \frac{|C_R-C_*|}{C_*}$. The credibility between $C_L$ and $C_R$ is $d =  \frac{|C_L-C_R|}{C_R}$. 
 Then the performance guarantee of $C_L$ could be estimated according to the following theorem. 
 
 \textcolor{black}{Then we have a theorem as follows:}  

\begin{theorem}
    If $\forall task$, $\varepsilon (n) \le \epsilon $ \& $d$ is a small constant,  then $\frac{|C_L-C_*|}{C_*} \le d(1+\epsilon) + \epsilon$
\end{theorem}

\emph{proof:} We first consider the minimization problem. The bound of our methodology could be defined as follows:

\begin{equation}
    \begin{aligned}
    \frac{|C_L-C_*|}{C_*} & = \frac{C_L-C_*}{C_*} \\
    &= \frac{(C_L-C_R) + (C_R- C_*)}{C_*} \\
    & = \frac{C_L-C_R}{C_R-C_*} \times \frac{C_R-C_*}{C_*} + \frac{C_R- C_*}{C_*} \\
    as \quad C_* &=\frac{C_R}{1+\varepsilon (n)} \\
    \frac{|C_L-C_*|}{C_*} & =\varepsilon (n) \times (\frac{C_L-C_R}{C_R} \times \frac{1+ \varepsilon (n)}{\varepsilon (n)} + 1)  \\
     & \le \epsilon \times ( d \times \frac{1+\epsilon}{\epsilon} + 1 )   \\
    & = d(1+\epsilon) + \epsilon
    \end{aligned}
\end{equation}

Similarly, there also exists a bound for the maximization problem. The SEA is proved be solid with some conditions.$\Box$

\section{Experience Enhanced Deep Learning}
\label{eedl}
In this section, we apply our methodologies to enhance the deep learning and design a novel deep learning, called experience enhanced deep learning(EEDL) for automatic database management. 
We first briefly introduce EEDL in Section~\ref{aeedl}. Then we introduce the training algorithm of EEDL in Section~\ref{eedltrain}.

\subsection{The Introduction of EEDL}\label{aeedl}

\begin{figure}[htb]
    \includegraphics[width=1\linewidth]{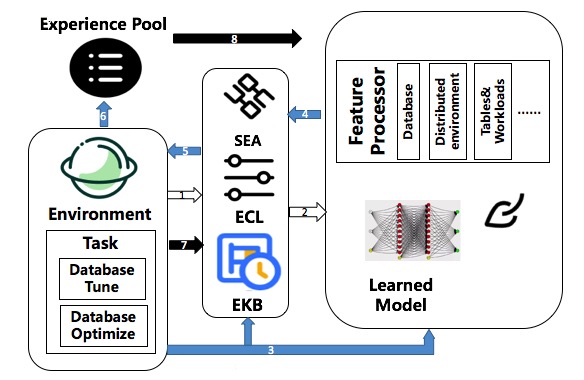}
    \caption{The Architecture of Experience Enhanced Deep Learning for Automatic Database Management}
    \label{fig:eedl}
\end{figure}

Deep learning is a black box which requires many samples to train and has unstable performance. Thus, we consider to improve deep learning from three aspects. Firstly, we employ the ELC to accelerate the label collection for fast training. Then we integrate the EKB to help deep learning model for efficiently transfer learning. Finally, we use the SEA to support a solid theoretical guarantee for the learned solution.
As shown in Figure~\ref{fig:eedl}, EEDL contains three important pipelines. The preparation phase(Steps 1-2) is responsible for pre-training of the learning model in Section~\ref{pre}. The \textbf{online phase}(Steps 3-6) presents the pipeline of model inference in Section~\ref{online}. The \textbf{re-train phase} (Steps 7-8) is in charge of model re-training and EKB updating in Section~\ref{retrain}.

\subsubsection{Preparation Phase}\label{pre}

All the deep learning models require extensive labeled data before they could work for database management tasks~\cite{neo}. The key sight of \textbf{preparation phase} is to efficiently obtain the training dataset. Due to the complexity of database system, it is time consuming to directly obtain enough pre-training samples by execution queries. In this phase, we utilize the Algorithm~\ref{alg:label} to accelerate training process. As shown in Figure~\ref{fig:eedl}, steps 1-2 represent the model pre-training. In this stage, EEDL gathers enough labeled data and delivers them to the learned model. 

\subsubsection{Online Phase}\label{online}

In this section, we introduce how to implement model inference. Model inference is a process that generates the required strategy of input task, such as the knob values of knob tuning and the cardinalities of query estimation.

For model inference, we integrate the Algorithm~\ref{alg:theore} to guarantee the stable performance of models. When processing a new task, EEDL first calculates a credibility about whether the learned model could handle this task. As we can observe in Figure~\ref{fig:eedl}, steps 3-6 is the process of model inference. EEDL firstly delivers a new task to `EKB' and `Learned Model'. We could obtain two solutions, $C_R$ and $C_L$. Then, EEDL uses them to count a credibility which represents the applicability of current model for this task. According to the credibility, EEDL chooses to utilize the rules or learning models to perform this task in step 5. After executing this task, EEDL could obtain a unique experience for current environment and stores this experiences to `Experience Pool'. These experiences are vital for model re-training, since they have completely unique labels to the current environment.

\subsubsection{Re-train Phase}\label{retrain}.
Re-train Phase is charge of in charge of model re-training and EKB updating. 

On the one hand, EDL could efficiently obtain a weak learned model from the preparation stage. However, such initial model only has an average performance of current environment due to the weak labeled data. And we use model re-training to improve the learned model. We use the methodology in Section~\ref{know} and construct an experience pool to store the practical experience from current practical environment. As shown in Figure~\ref{fig:eedl}, this phase contains two steps(steps 7-8). The first one is to sample a mini-batch data for model re-training. The sample method should be selected by a specific database management task, such as random sampling, biased sampling and so on. For a stable environment, we could use random sample to balance the new experiences and old once. For changing environment, we could sample more new experiences to adapt to the environment.

On the other hand, with the changes of environments, the selected rule-based methods of EKB may be improper to the task. For example, with the development of applications, the number of table changes from 1 to 10. The cardinality estimation method which is originally selected may not be applicable for 10 tables. Therefore, in order to ensure the adaptation of rule-based methods, after regular interval which is defined according to user demand, EEDL needs to evaluate the matching degree between current task and the selected rule-based methods by using the evaluation method introduced in Section~\ref{know}. If not proper, EEDL will search another proper rule-based methods for current environment.

\subsection{Training Algorithm}\label{eedltrain}
\textcolor{black}{The training algorithm of EEDL integrates the ELC method for fast model pre-training, utilizes the EKB to adapt to changing environments and employs the SEA to guarantee the stability of the learned solution.}
We show the training algorithm of EEDL in algorithm~\ref{alg:eedl}. The input of this algorithm consists of four parameters. $T$ is the queue of real-time task. $I$ is the interval condition of re-training which balance the adjustment cost and performance improvement. $d$ is the bound of credibility. $EKB$ is the knowledge base of automatic database management. This algorithm runs along the entire life cycle of the system for ensuring the high suitability of real environment. Next, we introduce this algorithm in detail.

As shown in Algorithm~\ref{alg:eedl}, Lines 1 and 2 init the set of practical experience $LS$ and the parameters of the model, respectively. Line 3-4 completes the pre-training of model. Line 5 selects proper rule-based approaches from EKB. Lines 7-10 calculate the credibility of task $t$ by employing the Algorithm~\ref{alg:theore}. This operation ensures that EEDL can have average performance for new tasks and could achieve higher availability than existing learned methods which maybe produces uncertain results for new tasks. In Lines 11-13, the credibility $c$ satisfies the bound and executes the task with the learned model. Lines 14-15 express executes task with rule-based methods. Line 16 updates the practical experience pool. When reach the condition $I$, Lines 18 implements the re-training of the model by sampling from the LS. Line 19 updates the selected rule-based methods which help to implement model transfer.

\begin{algorithm}
    \SetAlgoLined
    \LinesNumbered
    \SetKwInOut{Input}{input}
    \Input{
        \ $T$ is a queue of real-time task\\
        \ $I$ is the condition of re-training\\
        \ $d$ is the bound of credibility \\
        \ $EKB$ is the knowadge base\\
    }

    \BlankLine
        \ $LS \leftarrow  \varnothing$\\
        \ initialize $\Theta$  // \textit{\textcolor{blue}{init the learning model}} \\
        \ $TS \leftarrow ELC(T)$ \\
        \ $\Theta \leftarrow $  \textit{preTrain( $\Theta$, $TS$)} \\
        \ $Rule() \leftarrow EKB$ \\
    
            \While{true}{
                \ t = $T$.pop() \\
                \ $C_R \leftarrow Rule(t)$ \\
                \ $C_L \leftarrow Model(t) $ \\
                \ $c \leftarrow$ \textit{Confidence($C_R, C_L$)} // \textit{\textcolor{blue}{calculate the credibility}} \\

                \uIf{c $\le d$ }{
                    \textit{Execute}(t, $C_L$) // \textit{\textcolor{blue}{execute t with learning model}}\\

                }
                \Else(){
                    \textit{Execute}(t, $C_R$) // \textit{\textcolor{blue}{execute t with rule-based methods}}\\

                }
                \ $LS \leftarrow LS \cup $ \textit{gatherLable()} // \textit{\textcolor{blue}{obtain real label after execution}} \\
           
                \If{reach I}{
                   
                    \ $\Theta \leftarrow $ \textit{reTrain}( $\Theta, LS$) //\textit{\textcolor{blue}{reach the re-training condition}} \\
                    \ $Rule() \leftarrow EKB$ //\textit{\textcolor{blue}{update the rule-based methods with environment}} \\
                }
            }

\caption{The Training Algorithm of EEDL}\label{alg:eedl}

\end{algorithm}

\section{The evaluation of EEDL: Cardinality Estimation}
\label{evaeedl}
In this section, we compare the performance between EEDL and general DL~\cite{mscn} with a essential database task, cardinality estimation. We first introduce how we configure the EEDL to achieve cardinality estimation. Then, we introduce the experimental setup, containing the hardware settings, dataset and metric. Finally, we show the evaluation results of EEDL from three aspects: training time efficiency, Q-error with re-training and the robustness of the changes of re-training interval.

\begin{figure*}[ht]
    \centering
    \begin{tabular}{@{}c@{}}
        \includegraphics[width=0.4\linewidth]{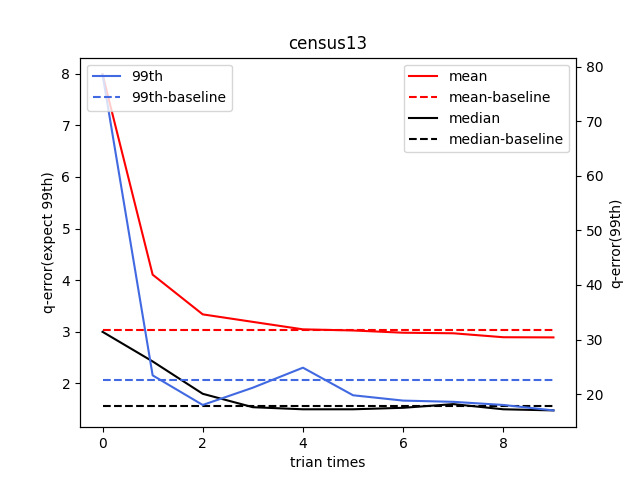}
        \includegraphics[width=0.4\linewidth]{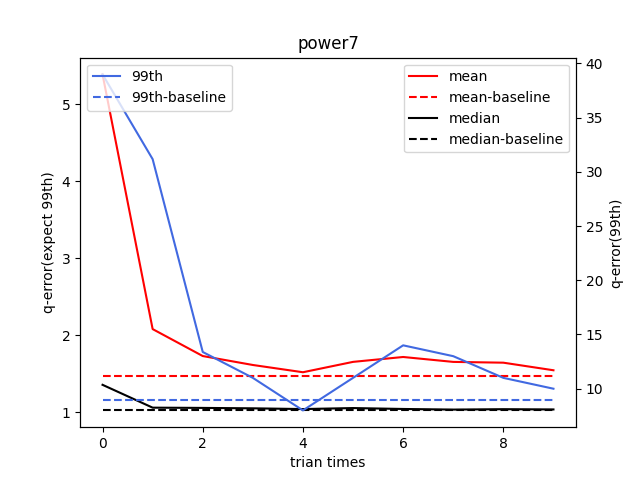}
    \end{tabular}
    \hfill
    \begin{tabular}{@{}c@{}}
        \includegraphics[width=0.4\linewidth]{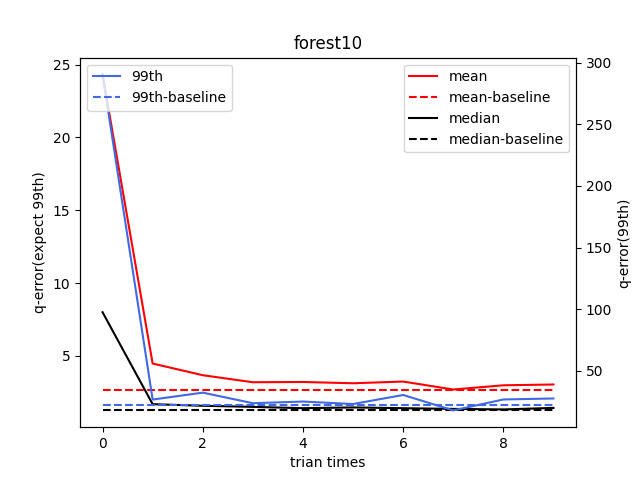}
        \includegraphics[width=0.4\linewidth]{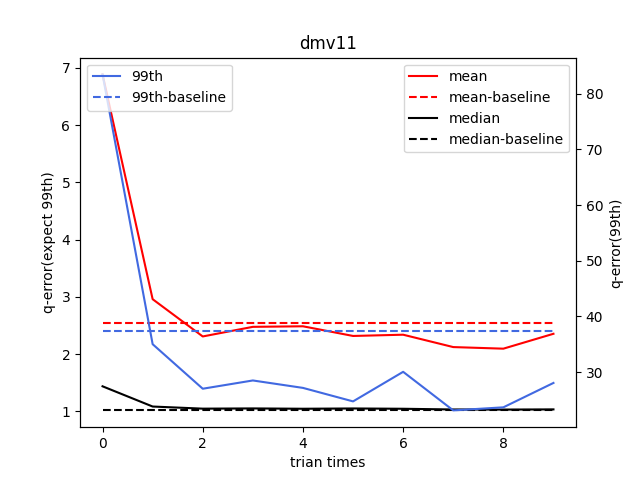}
    \end{tabular}
    \caption{The Q-errors of MSCN (dotted line) and the experience enhanced MSCN  (solid line) with the times of re-training in difference dataset.}
    \label{fig:rlreward}
    
\end{figure*}

\subsection{The Configuration of Cardinality Estimation}
\
For EEDL, we mainly need configure four factors with cardinality estimation, including EKB which integrates unified cardinality estimator, ECL which clarifies the label collection, learned model for cardinality estimation and the sample methods for re-training. The detailed information of configurations are as follows:

\begin{itemize}
    \item For cardinality estimation, we construct the EKB with a universal cardinality estimator in \textit{Postgres}~\footnote{https://www.postgresql.org} 11.6 as the experience rule methods, which could be used to labeling and transfer learning. 
    \item For configuring ECL, we first use the open source generation code~\cite{ready} designed by human as the method to generate many unlabeled queries in four dataset~\cite{census,dmv}. Then, we use the estimator of \textit{Postgres} as the rule-based methods to obtain weak labels for these queries. For every dataset, we generate 20K queries to implement the model pre-training. 
    \item We use the existing work MSCN~\cite{mscn} as the learned model, which is carefully designed for the cardinality estimation.
    \item For the sampling method, we choose the simple random approach in the `Practical Pool'. In the experimental result, we find that even we use the random sampling, there also be significant performance improvements, because the experiences in practical pool is unique for current environment.
\end{itemize}

\subsection{Environment Setup}
\

\textbf{Hardware and Platform:}
Our experiments conduct in Windows 10 system by using NVIDIA GeForce RTX 2060 GPU, 32 GB memory and Intel Core i7-10857H CPU.

\textbf{Dataset:}We employ the benchmark generated in the survey~\cite{ready}, contains four real-world dataset, Census~\cite{census}, Forest~\cite{census}, Power~\cite{census} and DMV~\cite{dmv}. These datasets have difference size (4.8MB, 44.3MB, 110.8MB, 972.8MB) and distribution in columns and workloads. Also, these datasets are used to test at least one existing learned cardinality method~\cite{ready}. In our experiment, we choose 20K queries as the preparation training samples and use the remaining queries as the execution query queue of testing online phase.

\textbf{Metric:}
We choose the widely used metric, q-error~\cite{card1, card2, card3} to measure our methods. Q-error is a ratio between the max and the min in estimation cardinality and actual cardinality. The formal formula is as follows:
\begin{equation}
    q-error = \frac{max\{estimation(q), actual(q)\}}{min\{estimation(q), actual(q)\}}
\end{equation}

\begin{figure*}[htp]
    \centering
        \includegraphics[height=3cm,width=11cm]{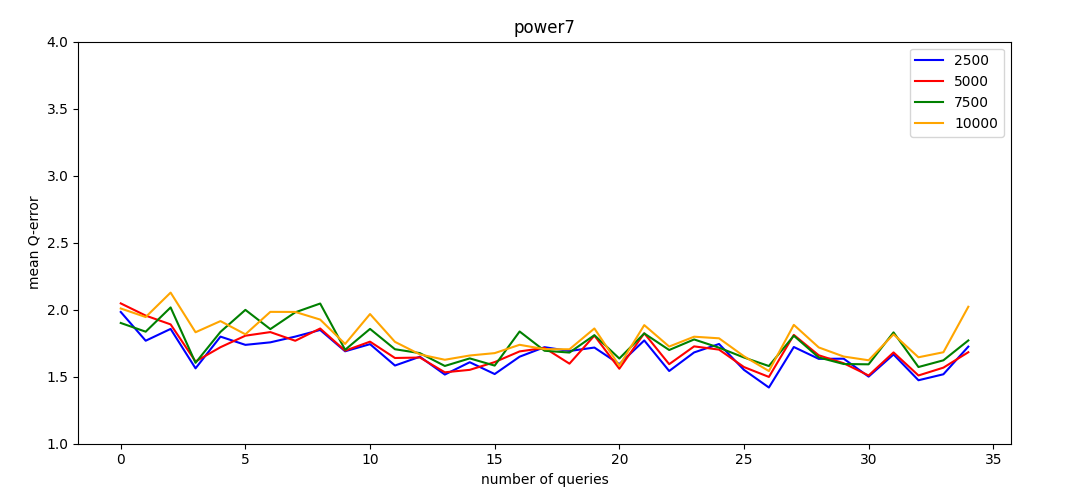}
            
    \caption{The mean Q-errors of re-training interval = 2.5K, 5K, 7.5K, 10K.}
    \label{fig:interval}
    
\end{figure*}

\subsection{Time Efficiency}
\
Firstly, we evaluate the time consumption of preparation phase between EEDL and general DL. As shown in Table~\ref{tab:eedl}, we test the time consumption of training process in four datasets. We can find that our EEDL which utilize the Algorithm~\ref{alg:label} to obtain labels only cost some seconds to get this labels, such as census13\: 3.4s, forest10:3.49, power7: 3.95s and dmv11:4.20s. While the general learned model which need to obtain labels from query execution takes more than 8 hours to gather cardinality labels in all dataset. EEDL could largely reduce the time consumption in model pre-training. And if we need to implement the learned model to real-world applications, the EEDL could quickly establish an average model by embedding the rule-based methods. 

\begin{table}[h]
    \begin{tabular}{||c c c||}
        \hline
            Dataset  & EEDL model (second) & general learned model\\
            \hline \hline
            census13  & 3.4048 & BAN \\
            \hline 
            power7    & 3.9522 & BAN \\
            \hline 
            forest10  & 3.4965 & BAN \\
            \hline 
            dmv11  & 4.1986 & BAN \\
        \hline
    \end{tabular}
    \caption{The time consumption in preparation phase (BAN presents that model could not obtain the labels within 8 hours).}
    \label{tab:eedl}
\end{table}

\subsection{Q-error with Re-training Times}

In this section, we test the adaptation of EEDL which employs the methodologies introduced in Section~\ref{methods}. We evaluate the Q-error changes with the times of re-training. For fair comparison, in all dataset, we implement the re-training efficiency with interval = 10K. And we test the Q-error with 2K queries in every re-training and MSCN. As shown in Figure~\ref{fig:rlreward}, we test the Q-error changes in four datasets and compare them to the MSCN~\cite{mscn}. The dotted lines represent the performance of general MSCN. And the solid lines represent the performance of EEDL based MSCN. We observe that after the preparation phase, the performance of EEDL based MSCN is totally poor on all indicators (99th, median, mean). And with the implementation of re-training, learned model start to learn the unique experience from the current environment. In all dataset, after two re-training, EEDL based MSCN start to steadily tend to the general MSCN. Specifically, in dmv11, our method outperforms the general MSCN on the 99th after the one re-training. That's because that our model could adapt to current environment by using the random samples from the `Experience Pool' in time while the MSCN pay more attention to the whole workload. From this section, we conclude that EEDL could achieve performance improvement with re-training.

\subsection{Q-error with Re-training Internal}

In this section, we test the robustness of the re-training interval parameter in dataset power7. As shown in Figure~\ref{fig:interval}, we evaluate the changes of mean Q-error with re-training interval = 2.5K, 5K, 7.5K, 10K. And we find that there is only 0.12 higher on average between the performance with 2.5K and the 10K interval. So, an appropriate interval is enough and we do not need to set the interval as small as possible.

\section{Experience Enhanced Reinforcement Learning}
\label{eerl}

In this section, we introduce the implementation of our methodologies in reinforcement learning. We design an experience-enhanced reinforcement learning(EERL) for automatic database management. We show the introduction of EERL in Section~\ref{aeerl} and clarify the training algorithm of EERL in Section~\ref{geerl}.

\subsection{The Introduction of EERL}\label{aeerl}
Different from the deep learning, reinforcement learning(RL) methods are active and unsupervised. The key sight of general RL is to learn from explorations, including random explorations and agent explorations. Thus, we cannot use SEA to fix the exploration result, which may influence the activity of RL. 
In this paper, we consider improving RL with rules from two aspects by integrating the methodologies introduced in Section~\ref{know}. One is to reduce the time consumption of agent training by integrating ECL. The other is to help RL to efficiently adapt the current environment by exploiting EKB. Although RL has the ability of transfer learning, the proposed EKB could enhance this process by accelerating the adaptation speed of agent.

\begin{figure}[htb]
    \includegraphics[width=1\linewidth]{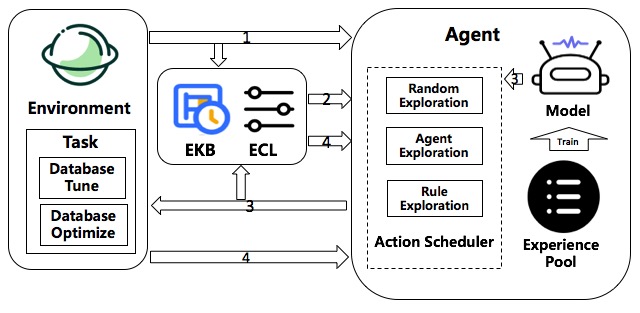}
    \caption{The Architecture of Experience Enhanced Reinforcement Learning for Automatic Database Management}
    \label{fig:eerl}
\end{figure}

Figure~\ref{fig:eerl} shows the pipeline of EERL, containing four main steps. Firstly, EERL delivers the current state to `Agent' and `EKB'. Then, for this state, both learned model and rule-based methods will produce an exploration suggestion, called rule exploration and agent exploration. Next, EERL decides the final direction of exploration by the `Action Scheduler'. After determining the direction of exploration, EERL obtains the reward label from ECL or real environment. Obtaining the reward label from ECL could significantly reduce the training time without query execution. Importantly, ECL could be used to obtain a batch labels and also has the ability to obtain the single label as the reward of RL. Obtaining reward from real environment could gather an accuracy reward for agent training but requires some time consumption. Thus, utilizing A or B depends on how expensive the reward is. 

Next, we further explain our key idea of controlling the direction of exploration. As shown in Figure~\ref{fig:eerl}, the `Action Scheduler' contains three directions including `Random Exploration', `Agent Exploration' and `Rule Exploration'. EERL could balance the convergence speed and solution quality by adjusting the rate of the above three explorations.

Compared to general RL, our EERL adds a parameter to control the rule explorations. For EERL, we design two hyper parameters ($\alpha, \beta$) and an attenuation plan to control the rate changes. $\alpha$ controls the rate of `Rule Exploration' while $\beta$ controls the rate of `Random Exploration'. Thus, $1-\alpha-\beta$ is the rate of `Agent Exploration'. The attenuation plan decides the changes of the exploration rate with iterations. The attenuation strategy is not necessary, and the users can use the fix rate to explore. 
Next, we introduce the three kinds explorations of EERL as follows.

\textbf{Rule Exploration:}
`Rule Exploration' means exploring the solution space by guiding with the rule-based methods, which has an average solution for current task. Rule-based guiding has two benefits.

On the one hand, the rules could efficiently guide the agent to an average performance and largely reduce useless explorations caused by the large search space. 
On the other hand, guiding by rules could improve the availability of the early reinforcement learning. In the early stage, the agent of reinforcement learning model performs very poor due to no experience and may conduct performance degrade in database system. Sometimes, for commercial database, the performance degrade would reduce the economic benefits. 

\textbf{Random Exploration:}
Although agent could quickly achieve convergence for current management task by using the `Rule Exploration', we also need to use the `Random Exploration' to explore whether there exists better management solution. However, due to the uncertainty of random exploration, this step may also lead to performance degrade. 
Thus, users have to carefully configure the $\beta$ to avoid performance degrade according to specific task.

\textbf{Agent Exploration:}
Different from the other two explorations, `Agent Exploration' is the real representation of model performance. And the result of `Agent Exploration' is guided by the estimation of every optional action. The agent would choose the best one to implement. In the early stage, the estimation has poor accuracy, so the choice of agent tends to random selection. After many iterations, the agent could make more stable choice by learning from explorations. 

In conclusion, EKB could help agent to efficiently adapt to current environment by reducing the number of explorations. Moreover, EERL improves the agent performance in the early stage by utilizing EKB.

\subsection{Training Algorithm}\label{geerl}

Since we add a novel direction of exploration, our EERL adds a parameter to control the exploration rate of `Rule Exploration', which could accelerate the agent convergence and achieve stable performance. We show our training algorithm of EERL in the Algorithm~\ref{alg:eerl}. The input of our algorithm consists of six parameters, stop condition($I$), environment($Env$), the rate of rule exploration ($\alpha$), the rate of random exploration($\beta$), the knowledge base of certain task ($EKB$) and the condition of updating $EKB$ ($W$). 
Next, we introduce the training algorithm of EERL in detail.

Firstly, Line 1 initializes the parameters of RL agent. Line 2 creates an empty experience pool $D$. Line 3 chooses the rule-based method from EKB. After obtaining the state of current environment(Line 4), EERL starts to repeat the following steps until satisfying stop condition $I$. Lines 6-8 calculate three actions from three explorations, i.e. $a_1, a_2, a_3$.
Line 9 samples the final action $a$ with certain probability. Then EERL updates the sample probability with the times of iteration $i$ (Line 10). Line 11 obtains the reward ($r$) and next state ($s'$) by executing the action in the environment or based on our ELC. It depends on how expensive the reward is. If we want to quickly complete the training process, we could use ELC. Line 12 stores this experience $\{s,a,r,s'\}$ to $D$. Line 13 samples a mini-batch from the experience pool to train agent. Line 15 judges whether the similarity of tasks and rule-based approach satisfies the update condition of $EKB$. If so, Line 16 implements the update operation.

\begin{algorithm}
    \SetAlgoLined
    \LinesNumbered
    \SetKwInOut{Input}{input}
    \Input{
        \ $I$ is the stop condition \\
        \ $Env$ is the environment\\
        \ $\alpha$ is the initial rate of rule exploration\\
        \ $\beta$ is the initial rate of random exploration\\
        \ $EKB$ is the knowledge base \\
        \ $W$ is the condition of updating $EKB$ \\
       
    }
 
    \BlankLine
    \ Init $\Theta$ \qquad  // \textit{\textcolor{blue}{the parameters of agent}}\\
    \ $D \leftarrow \varnothing$ \qquad  // \textit{\textcolor{blue}{built the experience pool}}\\
    \ $Rule() \leftarrow EKB$ \\
    \ $s \leftarrow Env.state()$ \qquad // \textit{\textcolor{blue}{obtain the current state}}\\

    \Repeat(){Satisfy I}{

        \ $a_1 \leftarrow Agent(\Theta, s)$  // \textit{\textcolor{blue}{calculate the action of agent}}\\
        \ $a_2 \leftarrow Rule(s)$ \qquad // \textit{\textcolor{blue}{obtain the rule action}}\\
        \ $a_3 \leftarrow Random(s)$ \qquad // \textit{\textcolor{blue}{obtain the random action}}\\

        \ $a \leftarrow Sample(a_1, a_2, a_3, \alpha, \beta)$  // \textit{\textcolor{blue}{sample action according to probability}}\\

        \ $update(i,\alpha, \beta)$ \quad // \textit{\textcolor{blue}{update probability with i}}\\

        \ $s', r \leftarrow Env.execute(a)$ | $ELC(s, a)$ \\
        \ $ D \leftarrow D \cup \{s, a, r, s'\}$ \\
        \ $\Theta \leftarrow  Agent.update(\Theta,D)$  // \textit{\textcolor{blue}{Sample mini-batch to train agent}}\\
        \ $ s \leftarrow s'$ \\
        \ \If{Satisfy W}{
            \ $Rule() \leftarrow EKB$ //\textit{\textcolor{blue}{update the rule-based methods with environment}} \\
        }

    }

\caption{The EERL Training Algorithm}\label{alg:eerl}

\end{algorithm}

\section{The evaluation of EERL: online index tuning} \label{evaeerl}
    In this section, we compare the EERL with the general RL in the online index tuning. We first introduce how to configure the EERL in online index tuning. Then, we introduce the experimental setup, containing the hardware settings, dataset and metric. Finally, we show the evaluation results of training time consumption, the convergence speed and Q-reward with different exploration rate.

\begin{figure*}[htp]
    \centering
    \begin{tabular}{@{}c@{}}
        \includegraphics[width=0.3\linewidth]{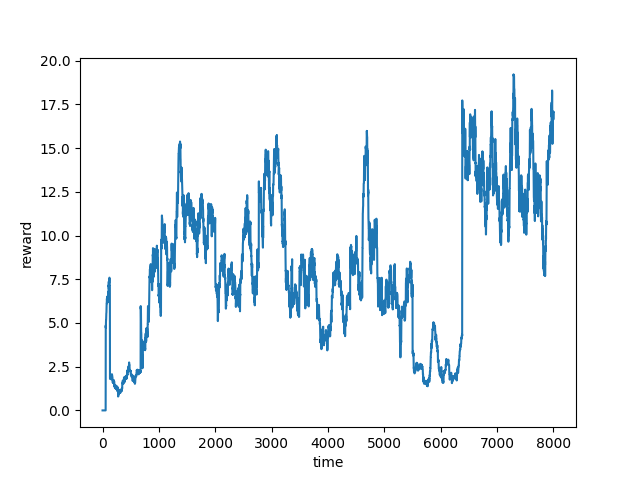}
    \end{tabular}
    \hfill
    \begin{tabular}{@{}c@{}}
        \includegraphics[width=0.3\linewidth]{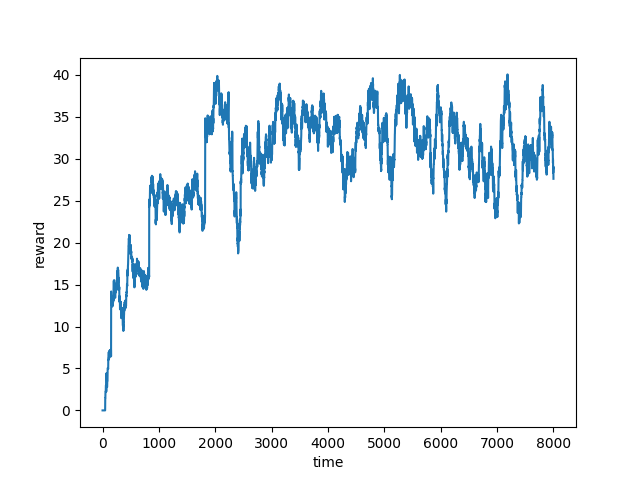}
    \end{tabular}
    \hfill
    \begin{tabular}{@{}c@{}}
        \includegraphics[width=0.3\linewidth]{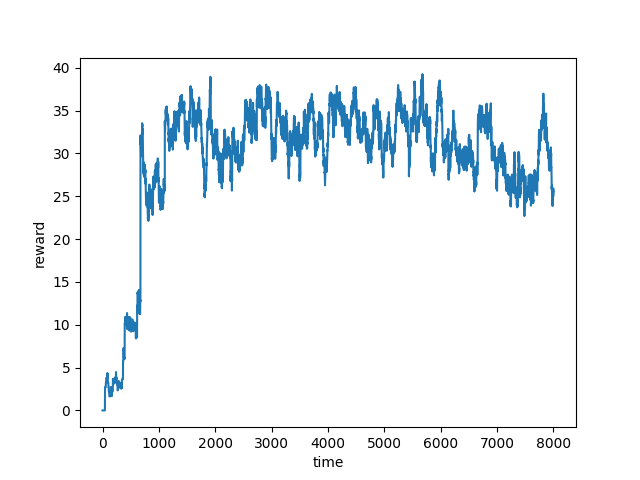}
    \end{tabular}
    
    \vspace{-1mm}
    \caption{The Q-cost of different initial rate. $a=0, b= 0.2$ (left figure) $a=0.2, b=0$ (medium figure) and $a = 0.1, b = 0.1$(right figure) with only 8000 iterations and no attenuation.}
    \label{fig:rlreward}
\end{figure*}

\subsection{The Configuration of Online Index Tuning}
\

Similar to EEDL, the configurations of EERL also mainly consists of four factors, including EKB which integrates unified rule-based index tuner, ECL which clarifies the label collection, learning agent for index tuning and the attenuation policy for the adjustment of exploration rate. We present the detailed information of configurations of online index tuning as follows:

\begin{itemize}
    \item For index tuning, we reference some existing rule based method~\cite{cophy,relax,recursive}. We observe that since index tuning is a complex task, the traditional rule-based method will take a lot of time to make decision. In order to facilitate the experiment, we construct the EKB with three universal rules based on the core idea(frequency and repetition) of existing works: (i)Choosing the frequent candidate to establish. (ii) Delete the repeated indexes. (iii) Delete the infrequent candidate. And experimental results in Section~\ref{res:eerl} shows that even if we use the above simple rules, EERL still outperforms general RL.
    \item ECL could be used to obtain a batch labels and also has the ability to obtain the single label as the reward of RL. For labeling index, we use the $what-if$ calls in Postgres as the rule-based methods which could establish the hypothetical indexes for obtaining the reward of indexes without index changes in real environment.
    \item For online index tuning, because there lacks details for reimplementing the existing works~\cite{reindex,onlineindex,nodba}, we realize a reinforcement learning model as the agent for online index tuning like these existing methods~\cite{reindex,onlineindex,nodba}. We use the workload and indexes as the input state as existing works\cite{nodba} do. Also, we use the index revision as the action set. 
    \item  For online index tuning, we employ a segmented attenuation strategy to ensure that learned model has difference attenuation ratio in different stages. The equation is defined as follows. $[0, c_1], [c_1, c_2], [c_2,]$ represent initial stage, learning stage, stable stage respectively. $\alpha$ is the initial rate and $iter$ is the number of iterations. For example, we define the initial rate of rule exploration $\alpha = 0.3$. And we want to utilize this rate before 3000 iterations. If we need to reduce this rate to 0.1 during 3000 to 5000 iterations, we could design $w = 0.2, c_1 = 3000, c_2 = 5000$. Then the rate of $\alpha$ will reduce from 0.3 to 0.1 after 5000 iterations.
    
\end{itemize}
\begin{equation}
    \alpha = \quad \left\{
                \begin{aligned}
                    & \alpha, &  iter < c_1. \\
                    & \alpha \pm w * \frac{c_2- iter}{c_2- c_1} , & c_1 \leq iter \geq c_2.\\
                     & \alpha \pm w , & iter > c_2   \\
                \end{aligned}
        \right.
    \end{equation}

\subsection{Environment Setup}
\

\textbf{Hardware and Platform:}
Our experiments conduct in Windows 10 system by using NVIDIA GeForce GTX 1050 Ti GPU, 16 GB memory and Intel(R) Core(TM) i7-7700HQ CPU.

\textbf{Dataset:} We use the widely used benchmark TPC-H~\footnote{http://www.tpc.org/tpch/} to test our EERL, which has eight tables. We utilize the 1GB scale and 18 query templates to construct the reinforcement learning environment of online index tuning. Based on the eight tables, we generate a query stream by utilizing the 18 query templates in TPC-H. 

\textbf{Metric:}
 We use the weighted sum of the query time improvement percentage (Q-reward) as the metric of index tuning. This metric could represent the query improvement in certain index. Our metric is defined as follows. $w_i$ is the weight of $q_i$. $ r_{i0}$ is the cost of $q_i$ without index. $ r_{ik}$ is the cost of $q_i$ with $index_k$. For fair comparison, we use 100 queries to calculate the Q-reward in the following experiments.
\begin{equation}
    Q-reward=w_1*r_{1k}/r_{10}+w_2*r_{2k}/r_{20}+...+w_n*r_{nk}/r_{n0}
\end{equation}

\subsection{Time Efficiency}
In this section, we examine the performance of ELC by evaluating the time cost of iterations. Unfortunately, actually performing indexing actions and measuring index returns in database are very time-consuming so that if we use the real reward, we could not get results within limit time. So, we only test the time consumption of 1000 iterations. We find that ELC only cost 3.68 minutes while execution in environment cost 1000 minutes. Our subsequent experiments are all based on ELC to obtain the reward value for training.

\subsection{Convergence Speed \& Q-reward}\label{res:eerl} 
\

In this section, we evaluate the convergence speed and Q-reward of the EERL which integrates the EKB to accelerate the convergence of agent. As shown in Figure~\ref{fig:rlreward}, we present the changes of Q-cost with the number of iterations. Next, we present some analysis for convergence speed and Q-reward.

 In order to clarify the convergence speed, we compare three situations with limited 8000 iterations. The left figure represents the general RL performance which has a fix probability of 0.2 for random exploration and a fix probability of 0.8 for agent exploration. We find general RL model has poor performance in convergence speed. Because the index search space for 8 tables is too large, and general RL may performs badly and requires many rounds of exploration to find good indexes. The medium figure presents the performance of only using `Rule Exploration'. We observe that `Rule Exploration' brings more stable performance improvements compared with other two situations. The last picture shows the situation in half of the random and the rule. We found that the EERL converges faster, which shows that random exploration also useful.

 For Q-reward, we observe that the left figure performs worse than other two results. Because in a big search space, the general RL with only random exploration may converge very slowly. And our EERL which integrates EKB could reach a good performance with limited iterations. 

\subsection{The influence of attenuation policy}

\begin{figure}[htp]
    \centering
    \begin{tabular}{@{}c@{}}
        \includegraphics[width=0.45\linewidth]{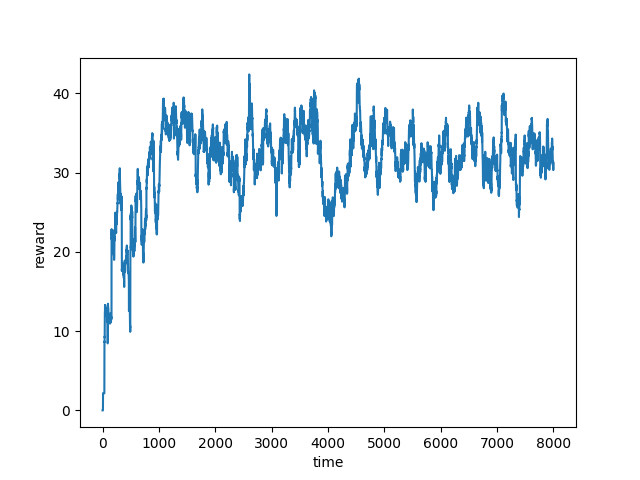}
        \includegraphics[width=0.45\linewidth]{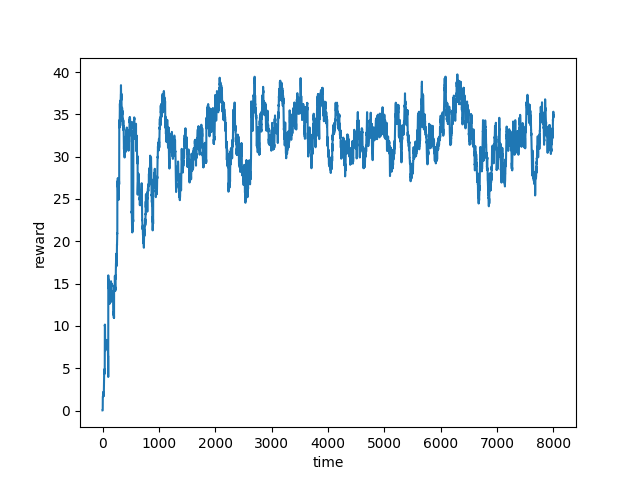}
    \end{tabular}
    \vspace{-1mm}
    \caption{The Q-cost of segmented attenuation policy. $a=0.8 \rightarrow 0.1 , b= 0.1 $ (left figure) $a=0.7 \rightarrow 0.1, b= 0.1 $ (right figure)) with only 8000 iterations.}
    \label{fig:rlatten}
\end{figure}

In EERL, the attenuation policy is also important which controls the rate of explorations. In this section, we present the evaluation results of attenuation policy. As shown in Figure~\ref{fig:rlatten}, utilizing the attenuation policy for $\alpha$ have more stable performance than the results in Figure~\ref{fig:rlreward}. Because the segmented attenuation could keep different exploration rate in difference stage. 

\section{Conclusion}
\label{conclusion}
    This paper tackles the challenges of deploying learned models to database management, containing \textbf{long training time, various environments and unstable performance}. We propose the methodologies which integrates the rule-based methods to enhance the learned methods, including rule-based label collection, knowledge base and theoretical guarantee methodology.  Specifically, we utilize our methodologies to two popular learning approaches, deep learning and reinforcement learning. The experimental results clarify that the learned models integrated with our methodologies outperform the individual learned models. Our methodologies also bring some new significant challenges for further research. For example, the construction of EKB has some challenges. Specifically, in each field of database management system, there exists many rule-based methods which could be used to enhance the learned model. It is a problem what standards we should use to unify them and construct the knowledge base, and how features to use to evaluate them with management tasks. In the future, we plan to explore more specific implementations of our methodologies.

\bibliographystyle{ieeetr}
\bibliography{citations/bibliography}

\end{document}